\documentclass[twocolumn,twoside,slac]{revtex4}
\usepackage{graphicx}
\usepackage{fancyhdr}
\begin{document}
\title{\hfill \parbox[t]{3cm}{\rm \normalsize DESY 03-073\\June 2003}
       \newline
       \newline
       Lattice QCD Calculations on Commodity Clusters at DESY}
\author{A.\,Gellrich, D.\,Pop, P.\,Wegner, H.\,Wittig}
\affiliation{Deutsches Elektronen-Synchrotron DESY,
             22603 Hamburg and 15735 Zeuthen, Germany\\
             (e-mail: Andreas.Gellrich@desy.de, Peter.Wegner@ifh.de)}
\author{M.\,Hasenbusch, K.\,Jansen}
\affiliation{John von Neumann Institut f\"ur Computing NIC and DESY,
             15735 Zeuthen, Germany\\
             (e-mail: Karl.Jansen@ifh.de)}
%
%
\begin{abstract} \label{abstract}
Lattice Gauge Theory is an integral part of particle physics that 
requires high performance computing in the multi-Tflops regime.
These requirements are motivated by the rich research program and the physics 
milestones to be reached by the lattice community.
Over the last years the enormous gains in processor performance, 
memory bandwidth, and external I/O bandwidth for parallel applications have made
commodity clusters exploiting PCs or workstations also suitable for large
Lattice Gauge Theory applications.
For more than one year two clusters have been operated at the two DESY sites
in Hamburg and Zeuthen, consisting of $32$ resp. $16$ dual-CPU PCs,
equipped with Intel Pentium 4 Xeon processors.
Interconnection of the nodes is done by way of Myrinet.
Linux was chosen as the operating system.
In the course of the projects benchmark programs for architectural studies were 
developed.
The performance of the Wilson-Dirac Operator (also in an even-odd preconditioned
version) as the inner loop of the 
Lattice QCD (LQCD) algorithms plays the most important role in 
classifying the hardware basis to be used.
Using the SIMD Streaming Extensions
(SSE/SSE2) on Intel's Pentium 4 Xeon CPUs give promising results for both the
single CPU and the parallel version.
The parallel performance, in addition to the CPU power and the
memory throughput, is nevertheless strongly influenced by the
behavior of hardware components like the PC chip-set and the
communication interfaces.
The paper starts by giving a short explanation about the physics 
background and the motivation for using PC clusters for Lattice QCD.
Subsequently, the concept, implementation, and operating experiences of the
two clusters are discussed.
Finally, the paper presents benchmark results and discusses comparisons
to systems with different hardware components including Myrinet-,
GigaBit-Ethernet-, and Infiniband-based interconnects.
\end{abstract}

\maketitle

\thispagestyle{fancy}

%
%
\section{Introduction} \label{introduction}
Lattice field theory has established itself as an integral part of
high energy physics by providing important, non-perturbatively
obtained results for many physical observables. It complements
standard approaches of theoretical particle physics such as 
perturbation theory and phenomenology,
becoming an indispensable method to allow for first principles
interpretation of experimentally obtained data. 
The aim of lattice field theory is to understand the 
structure of field theories, to test these theories against
experiment and in this way to find physics behind the standard model
of which we know that it has to be there, but not at what 
energy scale it should appear.
\par
Another important aspect of lattice field theory is its high 
computational needs.
A distinctive feature of the numerical computations in lattice field 
theory is the required performance of several Tflops and
this performance is needed ``in one piece''. Such a requirement can
only be fulfilled with massively parallel architectures having 
many processors that 
are connected via a very fast interconnecting network and
work simultaneously on the same problem as
a single machine. This distinguishes the computational needs in
lattice field theory from the farming concepts usually employed
in grid computing of experimental high energy physics.
\par
The combination of high computational needs and the motivation for
this in high energy physics as described above,
makes a conference such as CHEP an ideal
place to present the status and the perspectives of lattice field theory.
What strengthens the connection even more is that, at least,
a {\em data grid} is also needed in lattice field theory to exchange 
expensive data, the so-called configurations and propagators that are
generated.
An international lattice Data Grid initiative has been started \cite{ildg}.
\par
The computational requirements with high performance in the 
multi-Tflops regime and fine grained communication can be realized 
on commercial supercomputers like Hitachi \cite{hitachi} 
or IBM \cite{ibm}, with specialized 
machines such as APE (Array Processor Experiment) \cite{ape}
or QCDOC
(QCD on Chip) \cite{qcdoc}, or with PC clusters as they are discussed
in this article. 
Although commercial supercomputers
can conveniently be used, since they are maintained by the 
corresponding computer centers, their price is normally very high,
the efficiency of lattice field theory code is often not optimal and they
have many users from different application fields leading consequently
to a situation that a single user will not get much computer time.
\par
Specialized, custom-made machines like APE and QCDOC are cost-effective and
offer the best price/performance value for multi-Tflops installations. 
On the other hand, a lot of work has to be spent by the physicists themselves
in order to develop, build and maintain these machines. 
A compromise between supercomputers and custom-made machines might 
therefore be commercial PC clusters and in this
article\footnote{This paper covers three talks \cite{jansen2003, gellrich2003,
                 wegner2003} given during the parallel session on
                 {\em Lattice Gauge Computing} at
                 {\em Computing in High Energy Physics}, UCSD, La Jolla, USA,
                 March 24-28, 2003.}
we will concentrate on the experiences we have gained with these kind of 
machines at DESY. We also refer to \cite{evalgroup} for a
thorough discussion on the above point.
\par
One observation in lattice field theory 
is that people gather
in larger and larger collaborations. This results first of all from
the huge computer resources required. It is aimed to use these resources
wisely, not duplicating results and find the best strategies for
solving the physics problems.
One example of such an effort is SciDAC \cite{scidac}
in the US. Another one is the
Lattice Forum (LATFOR) initiative in Germany \cite{latfor}. 
Associated to LATFOR are also researchers in Austria and Switzerland.
In this initiative groups
from many universities and research institutes who work on lattice 
field theory combine their efforts to reach 
physics milestones in different areas of lattice field theory. 
They try to coordinate their physics program, to develop and share
software and share configurations and propagators, which play the
role of very expensive raw data for lattice field theory 
computations.
\newline
The research areas of LATFOR are broad and cover 
\begin{itemize}
\item ab initio calculations of QCD with {\em dynamical quarks}
     \begin{itemize}
     \item [ --  ] Hadron spectrum and structure functions
     \item [ --  ] fundamental parameters of QCD, i.e. the running 
     strong coupling $\alpha_s(\mu)$ and the quark masses 
     $\bar{\mathrm{m}}(\mu)$
     \item [ --  ] B-physics
     \end{itemize}
\item  matter under extreme conditions
     \begin{itemize}
     \item [ --  ] QCD thermodynamics
     \item [ --  ] QCD at non-vanishing baryon density
     \end{itemize}
\item  Non-QCD physics
     \begin{itemize}
     \item [ --  ] Electroweak standard model
     \item [ --  ] Supersymmetry
     \end{itemize}
\item  Conceptual developments                            
     \begin{itemize}
     \item [ --  ] exact chiral symmetry on the lattice
     \item [ --  ] acceleration of the continuum limit
     \item [ --  ] non-perturbative renormalization    
     \item [ --  ] finite size effects
     \item [ --  ] algorithm development
     \end{itemize}
\end{itemize}                                                     
It would be too demanding (and it is not the purpose of this
article) to discuss these topics in detail. The main target of
lattice calculations is certainly QCD and we will therefore give 
in the next section a --presumably too short-- introduction to
QCD and discuss a few examples of the results that can be obtained.
%
%
\section{Physics Motivation} \label{physics}
\subsection{Quantum Chromodynamics on the lattice}
In lattice field theory \cite{books}, the continuum 
space-time is replaced
by a $4$-dimensional, euclidean grid with a lattice spacing 
$a$, measured in fm. The advantage of this procedure is that 
now the theory can be simulated on a computer. 
In order to obtain back the desired results in the continuum, 
the values of observables obtained at non-vanishing values of the lattice
spacing have to be extrapolated in a continuum limit 
to zero lattice spacing.
At this stage,
the comparison to experiments becomes possible and a test of the
validity of the model considered can be performed. 
\par
On the pure computational side we are dealing with two kind of 
fields. The first one represents the quarks and are given by 
complex vectors
\begin{equation}
\Psi(x)_{\alpha,a,n_f},\;\; \left\{\begin{array}{ll}
                         \alpha=1,2,3 & \mathrm{color}\; \mathrm{index } \\
                           a=1,2,3,4  & \mathrm{Dirac}\; \mathrm{index } \\
                       n_f=1,\dots,6 & \mathrm{flavor}\; \mathrm{index }
                                \end{array} \right.\; .                          
\label{phidef}
\end{equation}
These fields live on the
$4$-dimensional space time point $x$ 
which, on the computer, is represented by integer numbers,
$x=(x_1,x_2,x_3,x_4)= a\cdot (i,j,k,l)\; ,
1\le i,j,k,l \le N$.  
A second set of fields represents the gluons of QCD and are
given by
SU(3) matrices $U$, $3\times 3$ complex matrix with unit norm.
The fields 
$U(x,\mu)_{\alpha,\beta}$ carry again color indices through which they
interact with the quark fields. The gluon fields live 
on the links of the lattice that connect points $x$ and $x+\mu$ in 
direction
$\mu=1,2,3,4$.                                     
The interaction is described by the
action\footnote{Of course, in QCD the quark fields are represented as Grassmann
variables. We discuss here the {\em bosonized} form of the action as it is used in simulations.}
\begin{equation}
S=\bar{\Psi}M^{-1}\Psi\; .
\label{action}
\end{equation}
\par
The action of eq.~(\ref{action}) requires the inverse of the 
so-called fermion matrix $M$,
or, to be more specific, the vector
$X=M^{-1}\Psi$. Without giving the exact definition of the 
matrix $M$, the problem is that the fermion matrix 
is high dimensional $O(10^6)\otimes O(10^6)$ and the 
numerical solution of the linear set of equations
\begin{equation}
M\cdot X=\Psi 
\label{inverse}
\end{equation}
employing such a matrix is clearly very
demanding. 
It helps, however, that the matrix is sparse. For such a case
a vast literature for solving eq.~(\ref{inverse}) exists
\cite{saad}. What is
important is that the algorithms that can be employed are 
self-correcting and many of them can be proven to converge in at most
a number of steps that corresponds to the dimension of 
the matrix. 
Thus we are
left with a well posed and regular numerical problem. 
Of course, in practice the number of iterations is 
much smaller and
typical numbers of iterations to solve eq.~(\ref{inverse})      
are $100 - 1000$. Note that in each of these iterations the matrix $M$
has to be applied to a vector of size $O(10^6)$.
\par
In order to give a feeling about the computational demand, let us
give an example. Let us consider a lattice of size
$32^3\cdot 64$ as is realistic in todays calculations in the 
quenched theory, where internal quark loops are neglected. 
Then we would need to solve eq.~(\ref{inverse}) 
twelve times per configuration for each color and Dirac component.
Each solution needs $O(200)$ iterations and we want to perform 
this on typically $O(1000)$ configurations. 
Since one application of the matrix $M$ on such a lattice needs
$2.8$\,Gflop, we are left with 
approximately $6.6$\,Pflop 
for obtaining only one physical result
for a single set of parameters, i.e. the bare coupling and the
bare quark
masses. 
On your standard PC which might run with $500$\,Mflops sustained
for this problem, you would hence need about five months. 
In order to reach control over the finite size effects,
the chiral extrapolation and the continuum limit, simulations at 
many values of the bare parameters have to be performed. 
\par
The numbers above hold for the quenched case, where the quark fields
are left out as dynamical degrees of freedom in the simulation.
If they are included, the cost of the simulations 
becomes at least a factor $100$ more and a single physical
result would need about $40$ years on your PC. Clearly, better computers
are needed such as the ones discussed in the introduction.
\par
Of course, relying only on progress in the development 
of computers would be too risky and not enough. Lattice
field theory has seen a number of conceptual improvements
\cite{oaimprove,chiral}
in the last years that allowed to accelerate the simulations
themselves. 
In addition, many improvements in the algorithms used were
found. Although each algorithmic improvement by itself
was only a relatively small step \cite{algoreview},
all in all a factor of $20$ acceleration through algorithm 
improvement alone could be obtained in the last $15$ years. 
Still, the development of machines were much faster in this
period. 
\begin{figure}[ht]
\begin{center}
\includegraphics[width=65mm]{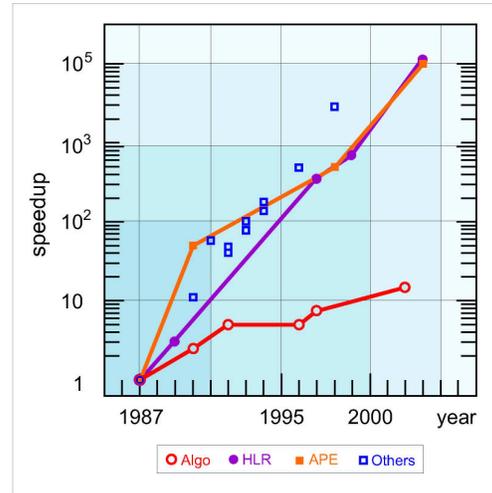}
\caption{\label{fig:speedup}
The performance gain of numerical simulations in lattice field theory
in the last years relative to the --normalized to one-- situation in 
1987. Algorithm development (Algo) alone reached a gain of a factor 
20. However, the performance gain through computer development appears 
to be orders of magnitude higher. 
We show this development at the example of the supercomputers 
exploited at the research center in J\"ulich (HLR) and of the APE
computers (APE). As a comparison we also show other architectures 
(others) as used worldwide for lattice field theory computations.
(taken from LATFOR).
}
\end{center}
\end{figure}
The situation is illustrated in 
fig.~\ref{fig:speedup} (taken from a LATFOR paper). Here we show
the speedup obtained relative to the status in the year 1987.
This year is special in that at that time the first exact algorithm 
for simulations of dynamical fermions was developed and used
\cite{hmc}. 
We see in the figure that since then the speedup resulting from better 
computer techniques, despite the impressive improvements 
by algorithmic developments, is orders of magnitude larger than the 
algorithm improvement. The figure also shows that special 
hardware, in this case the APE computer, shows the same scaling law as
commercial supercomputers, in this case the CRAY. 
Other computers and their performance,
relative to the status in 1987, used throughout the
world for lattice field theory is also shown. 
\subsection{Some selected physics results}
The machine, algorithmic and conceptual developments in lattice 
field theory allowed to compute a number of important physical 
quantities in the last years, despite the aforementioned very large
computational requirements. An important aspect of the 
developments is that we understand not only the statistical errors 
of the numerical calculations, but also the systematic ones. 
For a number of quantities fully non-perturbatively renormalized 
results {\em in the continuum limit} could be obtained. 
The only restriction of these calculations is that they are still done
in the quenched approximation. However, there is no particular 
reason why the calculation should not be doable in a completely
analogous way for the full theory. The advent of the next generation
of machines in the multi-Tflops regime will then open the door
to this exciting perspective.
\par
Let us just discuss two examples of physics results to illustrate 
what we just said. One is the running coupling \cite{rainerhart}
and the other are 
moments of parton distribution functions in deep inelastic
scattering. In a quantum field theory such as QCD, there is a steady
generation of virtual particles that shield (or anti-shield) the 
charges of the elementary particles. This leads to renormalization
effects. By changing the energy at which experiments are performed,
the scale at which we look at the, say, color charge is altered, and,
correspondingly the value of the charge itself depends on this
energy scale.
\par
This scale dependence (the running) 
of the coupling can be computed in lattice 
field, starting from the QCD Lagrangian alone, without any further 
assumptions. The trick is to use a suitable lattice renormalization
scheme, the so-called Schr\"odinger functional (SF) scheme,
that is defined in a finite volume and hence ideally suited 
for numerical simulations. By going to very high energies, contact
to perturbation theory can safely be established and renormalization
group invariant quantities can be extracted. In the case of the running 
strong coupling this corresponds to the $\Lambda$-parameter of 
QCD. 
The advantage of the knowledge of renormalization group invariant
quantities is that they can be translated to any preferred renormalization
scheme. In this way it becomes possible to translate results for
the running coupling from lattice simulations to continuum 
results in the, say, $\overline{\mathrm{MS}}$-scheme as it is 
conventionally used in perturbation theory.
\par
\begin{figure}[ht]
\begin{center}
\includegraphics[width=65mm]{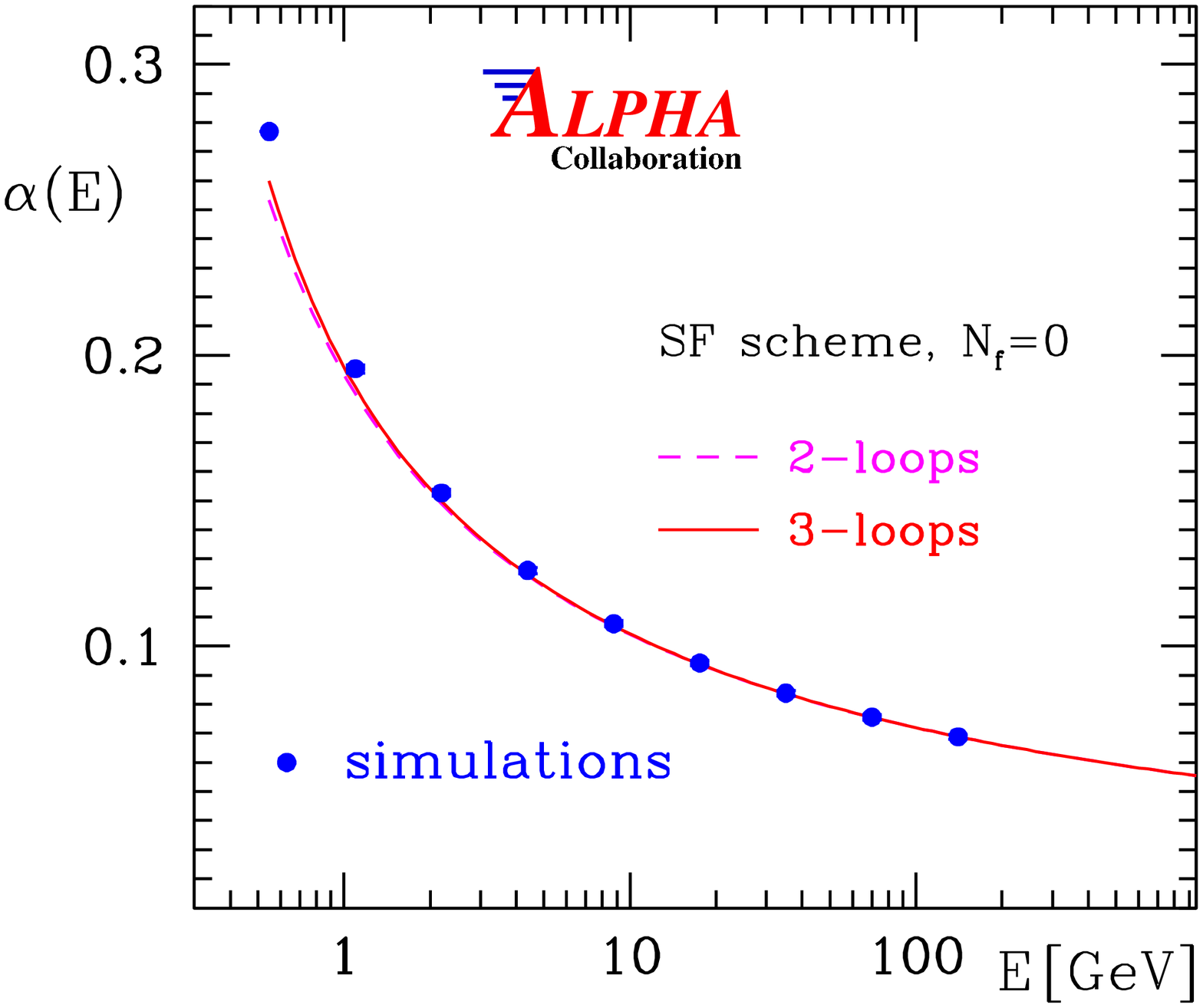}
\caption{\label{fig:runningcoupling}
The running strong coupling constant in the continuum as function
of the energy scale. The quenched approximation is used.
}
\end{center}
\end{figure}
\begin{figure}[ht]
\begin{center}
\includegraphics[width=65mm]{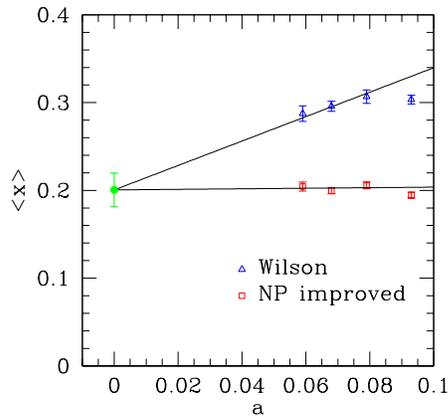}
\caption{\label{parton}
Continuum limit of the second moment of a twist-2 operator
in a pion. Two versions of lattice QCD are used, ordinary 
Wilson fermions (Wilson) and O(a)-improved fermions (NP improved). 
}
\end{center}
\end{figure}

In fig.~\ref{fig:runningcoupling} we show an example of such a calculation.
The results are already in the continuum. They cover a broad energy range
and are very precise (the error bars of the simulation points are
well below the size of the symbols). 
In the plot, also a comparison to perturbation theory is shown
and a good agreement down to surprisingly small energy scales
is found. 
Another example is a moment of a parton distribution function
as they can be extracted from global analyses of experimental
data. Such moments can be expressed as expectation values
of local operators and are hence computable in lattice 
simulations. The renormalization procedure of such moments follow
the general strategy of using the finite volume SF renormalization
scheme discussed above for the running coupling. 
We show in fig.~\ref{parton} an example of the continuum limit
of the first moment $\langle x\rangle$ 
of a twist-2, non-singlet operator in a pion
\cite{structuref}.
In the plot two different lattice formulations of QCD were used.
It is reassuring that in the limit that the lattice spacing is sent
to zero both unphysical lattice versions of QCD extrapolate to the
same number.
As a final result for this study it is found that the lattice gives
a preliminary value of $\langle x\rangle=0.30(3)$ while the experimental 
number reads $\langle x\rangle=0.23(3)$. This lattice number has to be 
taken with care since it is obtained in the quenched approximation.
But, again, there is no other reason than missing computer power 
to repeat this calculation also for the full theory. If in this case
the results just mentioned were stable, a really interesting situation
would have emerged. 
\par
The results that we have just discussed were actually obtained 
on APE machines. There are, however, a number of physics problems,
where PC clusters were used extensively.
The simulations concern in particular 
the recently discovered chirally
invariant formulations of QCD on the lattice.                    
Examples for results on the PC clusters that
are installed at DESY are given in \cite{clusterresults}.
The next sections are devoted to a discussion on PC cluster
systems that are installed at DESY. 
%
%
\section{Commodity Clusters at DESY} \label{system}
\subsection{Conceptional considerations}
Sufficient computing power to perform {\em Lattice QCD} (LQCD) calculations as
described above can obviously not be drawn from a single processor.
The solution is to parallelize the physical problem in order to concurrently
deploy many CPUs.
As a consequence massive intercommunication between the CPUs is needed.
\newline
Typical {\em High Performance Computing} (HPC) or {\em Supercomputing}
applications are characterized by high demands on:
\begin{itemize}
\item CPU (especially {\em Floating Point Unit} (FPU)) performance,
\item memory throughput,
\item interconnectivity (bandwidth and latency) between nodes.
\end{itemize}
One example for so-called {\em supercomputers} are
{\em Symmetric Multi-Processing} (SMP) machines with up to hundreds of
processors.
They appeared as single machines and are optimized for memory access and
interconnectivity between the CPUs.
\newline
In the LQCD area custom-made special purpose machines such as APE
(Array Processor Experiment) \cite{ape} or QCDOC (QCD on Chip)
\cite{qcdoc} have been developed.
\par
\begin{figure}[ht]
\centering
\includegraphics[width=65mm]{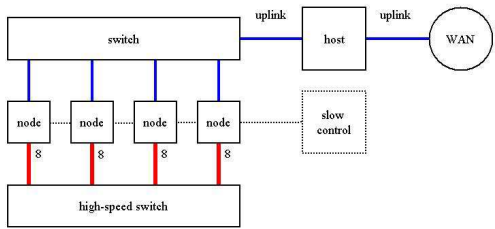}
\caption{Schematic view of a cluster.\label{fig:concept}}
\end{figure}
\begin{figure}[ht]
\centering
\includegraphics[width=65mm]{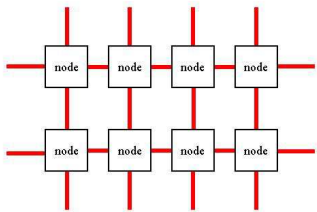}
\caption{Schematic view of a {\em mesh} cluster.\label{fig:concept2}}
\end{figure}

In the last years PCs, exploiting processors with competitive
computing power, hit the commodity market.
Concurrently, modern network technologies have achieved performances, which
allow for high-speed low latency interconnects between PCs.
These developments paved the way to build PC {\em clusters} \cite{clusters}.
\newline
Commodity clusters draw computing power from up to hundreds or even thousands
of in principle independent PCs with one or two CPUs, called {\em nodes}.
Those clusters benefit from their scalability and the possibility to deploy
components of the commodity market with good price/performance ratios.
Interconnectivity is provided by exploiting modern network technologies.
Such a cluster must:
\begin{itemize}
\item deliver sufficient CPU (FPU) performance and memory throughput,
\item provide good connectivity between CPUs (bandwidth as well as latency),
\item be scalable,
\item be reliable,
\item incorporate tools for easy installation and administration,
\item provide a usable software environment for the applications,
\item be connected to backup and archiving facilities,
\item fit boundary conditions such as space, cooling and power supply 
      capacities.
\end{itemize}
In clusters, a set of main building blocks can be identified:
\begin{itemize}
\item The {\bf computing nodes} which actually provide the computing power,
      optionally with local disk space,
\item a {\bf high-speed, low latency network} for the parallelized physics
      application,
\item an {\bf auxiliary network} to remotely control and administer the nodes,
\item a {\bf host system} for login, compiling, linking, batch job submission,
      and central disk space,
\item optionally, a {\bf slow control network}, e.g. based on a field bus.
\end{itemize}
A schematic view is shown in fig.~\ref{fig:concept}.
The high speed network can either be organized as a switched network
(e.g. the DESY clusters using  Myrinet-switches) or by a
n-dimensional {\em mesh} to allow for nearest-neighbor communication,
see fig.~\ref{fig:concept2}.
In \cite{Budapestcluster} the network is organized as a $2$-dimensional
GigaBit-Ethernet mesh.
\par
In scientific computing Unix-like operating systems have always played
the dominant role.
The development of Linux along with the triumphal procession of PCs
into the scientific world made Linux-PCs the systems of choice in most
universities, physics institutes, and laboratories.
In order to actually operate Linux-PC clusters, further system aspects must be
taken into account:
\begin{itemize}
\item Installation and administration of the {\bf operating system} Linux,
\item {\bf security} issues (login, open ports, private network),
\item user {\bf administration},
\item application software {\bf installation},
\item {\bf backup} and {\bf archiving} issues,
\item {\bf monitoring} and {\bf alarming}.
\end{itemize}
\subsection{Implementation}
The recently improved PC architectures are well suitable for Supercomputing
by exploiting:
\begin{itemize}
\item Increasing CPU clock rates following {\em Moore's Law} now extending the
      $2$\,GHz border,
\item larger caches at full processor speed,
\item vector units (SSE, SSE2),
\item cache pre-fetch,
\item fast memory interfaces,
\item PCI-Bus at $66$\,MHz/$64$-bit,
\item high external bandwidth (Myrinet, GigaBit-Ethernet), partly with very
      low latency (Myrinet).
\end{itemize}
\subsubsection{Hardware}
At both DESY sites in Hamburg and Zeuthen commodity clusters are operated
since January 2002 and December 2001 respectively.
See tab.~\ref{tab:clusters} for the set-ups.
\begin{table}[htb]
\begin{center}
\caption{DESY's PC clusters.\label{tab:clusters}}
\begin{tabular}{|l|l|l|}
\hline
\textbf{Item} & \textbf{Hamburg} & \textbf{Zeuthen}\\
\hline
Nodes           & $32$          & $16$          \\
CPUs/node       & $2$           & $2$           \\
CPUs/$1.7$\,GHZ & $2 \times 16$ & $2 \times 16$ \\ 
CPUs/$2.0$\,GHZ & $2 \times 16$ &               \\
\hline
\end{tabular}
\end{center}
\end{table}
\par
The cluster nodes as well as the host system are equipped with high-end
commodity components (see tab.~\ref{tab:nodes}).
\begin{table}[htb]
\begin{center}
\caption{Cluster Hardware.\label{tab:nodes}}
\begin{tabular}{|l|l|}
\hline
\textbf{Item} & \textbf{Implementation} \\
\hline
\multicolumn{2}{|c|}{Computing Nodes} \\
\hline
Chassis    & rack-mounted 4U module \\
Main-board & SuperMicro P4DC6 \\
Processors & $2$ Intel Pentium 4 Xeon $1.7$/$2.0$\,GHz \\
Chip-set   & Intel i860 \\
Memory     & $4 \times 256$\,MB RDRAM \\
Disk       & $18$\,GB SCSI IBM IC35L018UWD210-0 \\
\hline
\multicolumn{2}{|c|}{Host System} \\
\hline
Chassis    & rack-mounted 4U module \\
Main-board & SuperMicro P4DC6 \\
Processors & $2$ Intel Pentium 4 Xeon $1.7$\,GHz \\
Chip-set   & Intel i860 \\
Memory     & Rambus $4 \times 256$\,MB RDRAM \\
Disk       & $36$\,GB SCSI IBM DDYS-T36950N \\
Uplink     & Intel EtherExpress PRO 1000 F \\
Downlink   & Intel EtherExpress PRO 1000 T \\
\hline
\multicolumn{2}{|c|}{High-speed Network} \\
\hline
Interface cards & Myrinet M3F-PCI64B-2 \\
Chassis         & Myrinet M3-E32 5 slot \\
Line cards      & Myrinet M3-SW16-8F \\
Mngmnt card     & Myrinet M3-M \\
\hline
\multicolumn{2}{|c|}{Auxiliary Network} \\
\hline
Interface Card & on-board Intel 82557 100Base T \\
Switch         & Compu-Shack GIGALine 2024M \\
               & 48-port 100Base T \\
Uplink         & Module 1000Base T \\
\hline
\end{tabular}
\end{center}
\end{table}
\par
Intel Pentium~4 Xeon processors \cite{xeon}, which became available at the end
of 2001,
showed much enhanced performance compared to Pentium~III Tualatine CPUs due to
features such as vector units, and larger caches (see the
section~\ref{benchmarks} on benchmarking results).
The SuperMicro motherboard P4DC6 with the Intel i860 chip-set was the only
possible combination until the end of 2002.
It provides PCI-Bus support.
For the memory Rambus modules were chosen.
The communication between the nodes in the physics application is done by means
of Myrinet \cite{myrinet}.
It provides bandwidths up to $240$\,MB/s with very low latencies in the order
of a few $\mu$s.
For administration purposes and to actually submit jobs and copy data the nodes
are interconnected via Fast-Ethernet in a private subnet.
The host system is connected to the central switch by a copper GigaBit-Ethernet
link and has a separate fiber GigaBit-Ethernet link to the outside.
Each node consists of a PC with two CPUs and is housed by a 4U rack-mounted
chassis.
Up to $9$ nodes ($18$ CPUs) are installed in a cabinet.
\subsubsection{Software}
The basic installation plus most of the software support was purchased with
the hardware from the German company MEGWARE~\cite{megware}.
\newline
For all software related aspects the host system is used as a server for
the nodes.
\newline
For the Linux operating system a S.u.S.E. 7.2 distribution \cite{suse} was
chosen, which contains the kernel version 2.4.17 with SMP capabilities.
Temperature and fan sensors are read out by the kernel module {\em lmsensors}
\cite{lmsensors}.
\newline
The nodes are booted via {\em DHCP}, {\em TFTP}, and Intel's \cite{intel}
{\em Pre-boot Execution Environment} (PXE).
\newline
The nodes are operated in a private network (192.168.1.0) behind the server.
For security reasons,
external user login is only possible to the host system via {\em ssh}.
Individual login from the host system to the nodes can be done over the
auxiliary network by means of {\em rsh}.
Users are registered on the host system which exports the home directories
({\tt /home}) and a data partition ({\tt /data}) to the nodes.
For the user administration standard Unix tools are used ({\tt useradd}).
The necessary files ({\tt /etc/groups, /etc/passwd, /etc/shadow}) are
manually distributed to the nodes.
\newline
Source code is compiled and linked on the host system.
Software is mostly written in C/C++ but also compilers for Fortran77/90
are required.
On the host system in addition to GNU compilers of the Portland Group \cite{pgi},
KAI \cite{kai}, and Intel \cite{intel} are available.
\newline
The parallelization of the computation is done within the application by means of
the {\em Message Passing Interface} (MPI).
Since MPI is running over Myrinet, a special library which uses low level Myricom
communications is installed (MPICH-GM \cite{mpich-gm}).
For the Myrinet network a static routing table is used.
\newline
Zeuthen uses the open source {\em Portable Batch System} (OpenPBS) for
job submission.
In Hamburg nodes are manually distributed to users on good-will basis.
\newline
Time synchronization is done via {\em XNTP}.
The nodes synchronize with respect to the host system which gets the correct time
from DESY's central server.
\par
\begin{figure}[ht]
\centering
\includegraphics[width=65mm]{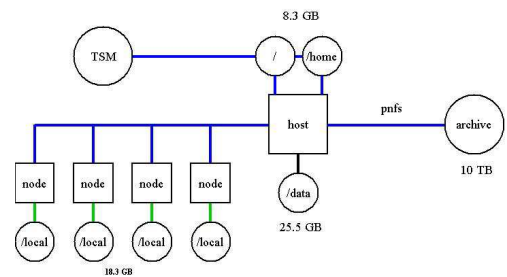}
\caption{The backup and archiving scheme.\label{fig:archive}}
\end{figure}

Backup and archiving are basically different items:
Regular backups are done to provide security against loss of system data and
home directories.
Under normal conditions backups will never be retrieved.
For the Hamburg cluster DESY's standard backup environment based on
IBM's {\em Tivoli Storage Manager} (TSM) is used.
It automatically creates incremental backups of the disk of the
host system and a regular basis.
At DESY Zeuthen a copy of the host system's disk is stored on a second
disk.
\newline
Archiving tools allow users to arbitrarily store and retrieve large amounts
of data.
DESY uses {\em dCache} which provides a
simplified and unified tool to access the tertiary storage \cite{dcache}.
It provides a unique view into the storage repository, hiding the physical
location of the file data, cached or tape only.
Dataset staging and disk space management is performed invisibly to the data
clients.
Currently around $1$\,TB of data are stored.
The archiving system is mainly used to store temporary {\em check-points} and
final results of long time computing jobs, so-called {\em configurations},
which can be used for further analyses.
\newline
The backup and archiving scheme is shown in fig.~\ref{fig:archive}.
\par
LQCD calculations require the availability of {\bf all} nodes dedicated to the
problem for the {\bf entire} run-time of the job.
The failure of only one node spoils the entire job.
Therefore, stability of the cluster in terms of availability and sustained
performance is a crucial credential of the clusters.
Usually not all nodes are used for one single job.
Many calculations are done on a limited number ($2-8$) nodes.
This allows to (manually) restart jobs using the check-points on different 
nodes in case of failures.
In order to use the resources of the clusters most efficiently a
well-defined monitoring and alarming scheme is needed.
\newline
The company MEGWARE delivered with the software installation a
monitoring package called {\em clustware}.
It provides a snapshot of all relevant properties of the cluster nodes in
a graphical user interface, including CPU usage, load, I/O, and temperature.
A long term history ($> 1 min$) is not shown.
\newline
Alternatively, a DESY in-house development called {\em ClusterMonitor}
(CluMon)~\cite{clumon} is in use.
Every node runs a simple Perl-written daemon which periodically dumps status
information of relevant node properties such as uptime, load, CPU usage, memory
usage, swap usage, and all temperatures into a node-specific file on the host
system.
The host system runs an {\em Apache} web server, which allows to remotely access
status information of the cluster with any web browser.
History of the quantities is kept by means of the {\em MRTG} package
\cite{mrtg} and is available from the web page.
In addition, {\em CluMon} provides alarming by e-mail based on the time period
since the last update of the status file. 
\subsection{Operational experiences}
The two clusters in Hamburg and Zeuthen are located in the
computer centers.
The clusters are operated and administrated in cooperation by members of the
computer centers and the DESY theory and NIC groups.
\newline
Of around twenty registered users per cluster just a handful can be
classified as {\em Power Users}, running regularly resource consuming jobs
(see tab.~\ref{tab:users}).
\begin{table}[ht]
\begin{center}
\caption{Accounts and users.\label{tab:users}}
\begin{tabular}{|l|l|l|l|}
\hline
\textbf{Site} & \textbf{Accounts} & \textbf{Power Users} & \textbf{Strategy} \\
\hline
Hamburg & $20$ & $7$ & good-will basis \\
Zeuthen & $18$ & $6$ & batch system    \\
\hline
\end{tabular}
\end{center}
\end{table}
\par
The two clusters in Hamburg and Zeuthen deploy in total $52$ dual-CPU PCs:
$32$ nodes (Hamburg),
plus $16$ nodes (Zeuthen), 
plus $2$ spare nodes,
plus $2$ servers.
During the almost $17$ month of operation quite a number of problems occurred,
so usually not {\bf all} nodes have been available at {\bf all} times.
Tab.~\ref{tab:errors} lists all major hardware problems.
\begin{table}[ht]
\begin{center}
\caption{List of failures.\label{tab:errors}}
\begin{tabular}{|l|l|l|l|}
\hline
\textbf{Component} & \textbf{Faults} & \textbf{Total} \\
\hline
\multicolumn{3}{|c|}{PC} \\
\hline
Motherboards    & $1$  & $52$ \\
CPUs            & $0$  & $104$ \\
Memory Modules  & $0$  & $208$ \\
Power Supplies  & $1$  & $52$ \\
Disks           & $35$ & $52$ \\
Ethernet Chips  & $0$  & $52$ \\
CPU Fans        & $0$  & $104$ \\
Chassis Fans    & $0$  & $52$ \\
\hline
\multicolumn{3}{|c|}{Myrinet} \\
\hline
Fibers          & $1$ & $48$ \\
Slot Chassis    & $0$ & $2$ \\
Line Cards      & $4$ & $10$ \\
Interface Cards & $6$ & $48$ \\
\hline
\multicolumn{3}{|c|}{Infrastructure} \\
\hline
Cabinet Fans & $3$ & $24$ \\
\hline
\end{tabular}
\end{center}
\end{table}
\par
Taking into account that the clusters exploit commodity hardware components
and offer considerably better price/performance ratios than big mainframe
SMP-machines, failures of certain components such as disks and power
supplies were expected.
The stability of the PC hardware after replacing the obviously systematically
misbehaving IBM disks was reasonable.
Some annoyance was caused the repeating failures of the Myrinet interface and
line cards, which was also seen at Fermilab \cite{holmgren} and traced back to
broken optical receivers.
Nevertheless, the general opinion of the users on the performance and the 
stability of the clusters is very positive.
\subsection{Future developments}
The considerations in section~\ref{physics} require cluster sizes of $O(1000)$ 
nodes to approach the Tflops regime.
Even more, in order to actually deliver a few Tflops sustained for hours,
days or even weeks, {\bf all} nodes would need to run at the same time.
As discussed earlier the failure of just one node would spoil the {\bf entire}
calculation.
Accepting the experiences so far, this seems not be possible within the current
concept\footnote{$1$ broken node per week in a $50$ node cluster is equivalent to
                 a maximal lifetime of a complete $1000$ node cluster of
                 $8$ hours.}. 
\newline
Recent tests have shown that GigaBit-Ethernet might be an interesting
alternative to Myrinet.
Benchmark showed bandwidths of $2 \times 1$\,Gbit/s bidirectional with special
switches which would imply a much better price/performance ratio than Myrinet.
Since GigaBit-Ethernet is widely used now, one might also expect more stability
and reliability compared to the niche product Myrinet
(see section \ref{benchmarks}).
\newline
Disks --even SCSI-- are the most likely components to break in PCs.
Though the replacement of disks is easy, the affected node is down and
needs to be re-installed completely afterwards.
This could be avoided by running the nodes disk-less.
Such a concept would require a stable and reliable server which could be achieved
by setting up one or more RAID-systems in a tree-like architecture to distribute 
load.
The server could also provide the boot image.
In another scenario booting could be done from a local block device such as an
EPROM or a memory stick.
\newline
The current set-up relies on one single server which exports {\tt /home} and
{\tt /data} directories to the nodes.
It also serves as a login host for the users and is used for code
development, compilation, linking, and job submission. 
This machine is clearly a single point of failure.
In a bigger system one would opt for redundancy in the server arrangement
by distributing different functionalities to different machines.
In particular a separate file server for the exported directories is needed.
\newline
Space, power consumption, and cooling will become a major issue when planning
for thousands of nodes.
Recent developments of so-called {\em blades} place the motherboards vertically
to improve the air-flow for cooling in order to increase CPU densities.
\newline
Software installation, administration, and monitoring of thousands of nodes
is a challenge which requires a very careful choice of appropriate tools.
Remote administration could be enabled by exploiting the serial consoles
of the PCs.
They could be connected to a dedicated terminal server or --in a much cheaper
scenario-- subsequently from node to node.
%
%
\section{Benchmarks} \label{benchmarks}
The hardware of commodity PCs has been extremely improved over the last 
years. The performance increase inside the  CPU is due to higher clock 
rates and enhanced building blocks (e.g. SSE1/SSE2 instructions) following
Moore's Law which predicts  a performance doubling all $18$ month. Moore's Law 
gives a technology estimation of mainly the CMOS density or number of 
transistors which can be integrated on a chip of a given size. It does not 
work well for the other interacting PC components like the memory interface 
and external busses. On the other hand also a big step forward in the 
development of fast memory architectures like Rambus and DDR RAM and
a series of high bandwidth PCI bus based interconnects like Myrinet and QSNet
is going on. Therefore PC clusters are becoming more and more attractive 
for classical high performance parallel computing and therefore also as a 
hardware basis for LQCD applications.
\subsection{Benchmark systems}
Apart from the DESY clusters described above, the following systems were used in 
order to test the ability of PC clusters for LQCD applications:\\
{\bf Mellanox}:  Blade dual Pentium~4 Xeon cluster connected via Infiniband, 
running MPICH for VIA/Infiniband with patch from Ohio State University
\cite{mellanox},\\
{\bf ParTec}: Dual Pentium~4 Xeon cluster connected via Myrinet running
ParaStation MPI~\cite{partec},\\
{\bf MEGWARE}: Dual Pentium~4 Xeon cluster connected via Myrinet running 
MPICH-GM from Myricom~\cite{mpich-gm},\\
{\bf Leibniz-Rechenzentrum Munich}: (single CPU tests) Pentium~4 and dual Xeon
PCs with CPUs with clock rates between $2.4$ and $3.06$\,GHz,\\
{\bf University of Erlangen}: GigaBit-Ethernet dual Pentium~4 Xeon cluster.\\
\subsection{Benchmarks and results}
\begin{figure}[ht]
\begin{center}
\includegraphics[width=65mm]{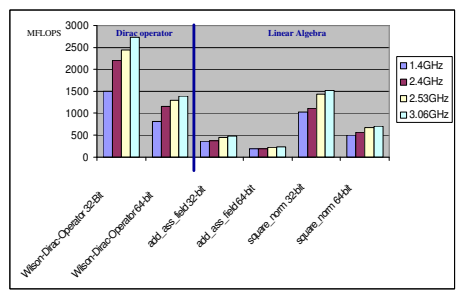}
\caption{Single node Wilson-Dirac operator and linear algebra benchmark.
         \label{fig:diagram1}}
\end{center}
\end{figure}

\begin{figure}[ht]
\begin{center}
\includegraphics[width=65mm]{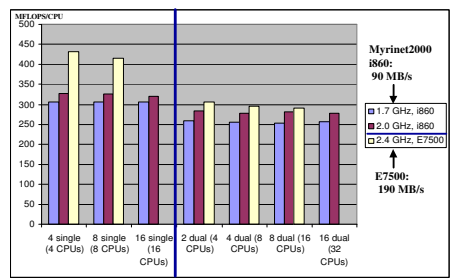}
\caption{Parallel ($1$-dim) Wilson-Dirac Operator Benchmark (SSE), even-odd 
         preconditioned, $2 \times 16^3$ lattice, Xeon CPUs, single CPU
         performance.\label{fig:diagram2}}
\end{center}
\end{figure}

\begin{figure}[ht]
\begin{center}
\includegraphics[width=65mm]{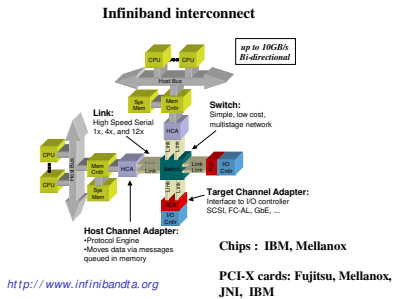}
\caption{Infiniband interconnect.\label{fig:diagram3}}
\end{center}
\end{figure}

\begin{figure}[ht]
\begin{center}
\includegraphics[width=65mm]{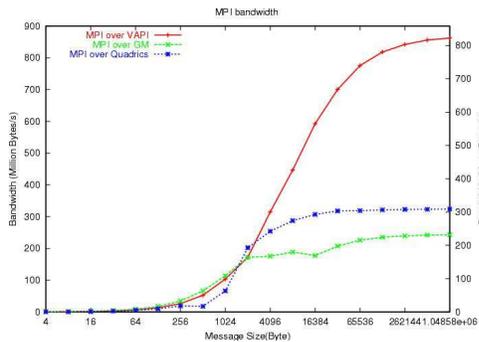}
\caption{Infiniband bandwidth compared to Myrinet and QSNet (source Mellanox).
         \label{fig:diagram4}}
\end{center}
\end{figure}

\begin{figure}[ht]
\begin{center}
\includegraphics[width=65mm]{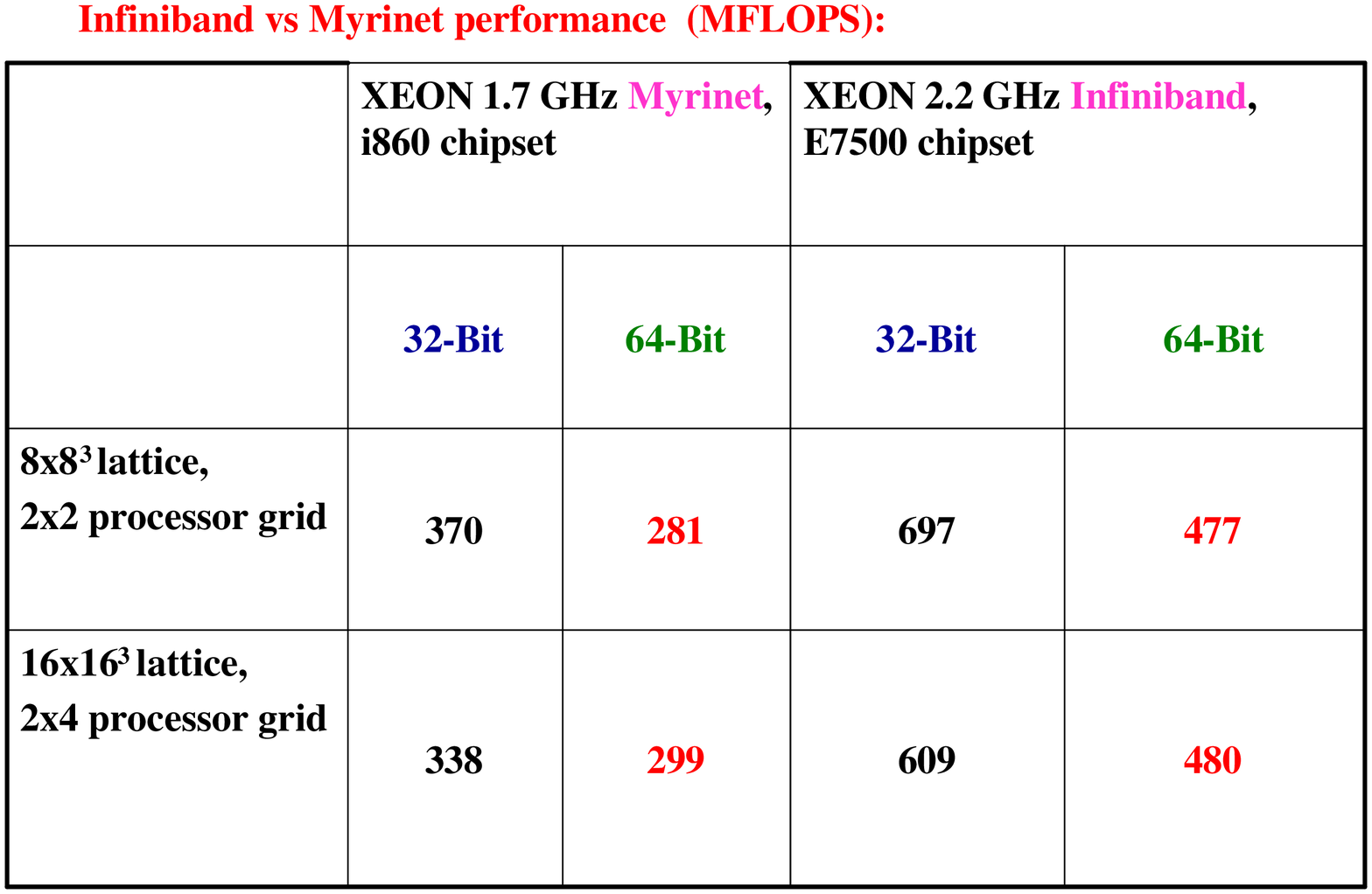}
\caption{Infiniband and Myrinet performance comparisons using a parallel 
         ($2$-dim) Wilson-Dirac Operator Benchmark on $4$ node Pentium~4 Xeon 
         clusters, single CPU performance, without SSE optimization, the local 
         lattice size is $4^2$x$8^8$ for the $8$x$8^3$, and $8$x$4$x$16^2$ for
         the global $16$x$16^3$ lattice.\label{fig:diagram5}}
\end{center}
\end{figure}

Representative benchmarks for the evaluation of different PC systems have 
been developed. Already in the year 2000 a first benchmark of M. L\"uscher 
(CERN) has shown the potential in using the SSE1 and SSE2 instructions for 
the Wilson-Dirac operator~\cite{Luscher2001}. This program takes heavily 
advantage of the Pentium~4 memory-to-cache pre-fetch capabilities and the 
SSE registers and instructions  which are implemented by using assembly in-line 
code, compatible to the gcc  and Intel compilers. Fig.~\ref{fig:diagram1} 
shows on the left hand side the performance gain of the highly optimized 
$32$-bit and $64$-bit Dirac-Operator kernel which linearly follows the 
evolution of the CPU performance expressed by their clock rate. The value 
of $1.5$\,Gflops for the $32$-bit implementation or 
$0.8$\,Gflops for the $64$-bit implementation respectively was unexpected high 
for a PC in the year 2000 and encouraged groups working on LQCD algorithms on 
PC clusters also to use the Pentium~4 capabilities to improve their algorithms 
on PC clusters.\\
The Wilson-Dirac operator Benchmarks are accompanied by two tests called 
add\_assign\_field (similar to the BLAS daxpy) and square\_norm which are
representing the linear algebra part of the benchmark. Both parts are strongly
memory bound which means that they cannot benefit from the SSE-environment.
This results in a relative small improvement shown on the right hand site of 
fig.~\ref{fig:diagram1} which also gives an impression of the slowly evolving 
memory interface architectures since the introduction of the dual channel 
$800$\,MHz Rambus.\\
Another version of such a single node benchmark was developed by M.\,Hasenbusch 
(DESY)
for the even-odd preconditioned Wilson-Dirac operator~\cite{cluster2002}.
Meanwhile (using recent 
FSB800 based PCs equipped with a $3.06$\,MHz Pentium~4) a performance of about 
$2.6$\,Gflops for the $32$-bit implementation and about $1.4$\,Gflops for the 
$64$-bit 
implementation can be observed.\\
To evaluate the behavior of PC cluster interconnects a $1$-dimensional parallel  
even-odd preconditioned Dirac Operator Benchmark on a $2 \times 16^3$ lattice 
(also written by M.\,Hasenbusch) was used (see Appendix). The aim of the parallel
benchmark was to compare different parallel PC based architectures against each 
other
rather than achieving the best performance for a given system.
Fig.~\ref{fig:diagram2} shows the results on clusters with different numbers of 
nodes.
In addition to the 
CPU power and the memory interface the throughput of the external PCI-Bus 
depending
on the given chip-set and the interconnecting interface itself are dominating 
the entire 
performance. Both early Intel Pentium~4 Xeon based clusters at DESY are using 
the i860 chip-set
which came with a relative poor $33$\,MHz/$64$-bit PCI-Bus performance.
A bus-read  (send) of
$227$\,MB/s and a bus-write (recv) of $315$\,MB/s of maximal $528$\,MB/s and
expected $450 - 460$\,MB/s was measured. This ends up in an external 
unidirectional
bandwidth of about $160$\,MB/s of maximal $240$\,MB/s. In the bidirectional case 
we measured $(90 + 90)$\,MB/s. Meanwhile more advanced chip-sets like the E7500 
were available for the Pentium~4 Xeon which are providing a PCI throughput close
to the expected 
numbers. The influence of the chip-set is dominating the results shown in
fig.~\ref{fig:diagram2}, whereas in the left hand side on CPU per node and an 
the right 
hand side two CPUs per node communicating via shared memory MPI are used.\\
Infiniband is a new promising communication technology especially designed for 
cluster 
interconnections fig.~\ref{fig:diagram3}. Beside the high throughput shown in 
fig.~\ref{fig:diagram4} the latency at relative small buffer size in the order 
of $2$\,kB is
significantly higher then using Myrinet which could be an advantage for 
applications
using small local lattices. In fig.~\ref{fig:diagram5} the performance of a $4$
node partition of
the DESY Myrinet Pentium~4 Xeon cluster is compared to the performance of a 
corresponding Infiniband
cluster from Mellanox  using a $2$-dim parallel Wilson-Dirac Operator Benchmark 
developed at DESY Hamburg by M. L\"uscher.
Due to some problems using the assembly in-lines within the 2.96 gcc compiler 
coming with
the RedHat Linux system on the Mellanox cluster during the short time in which 
the cluster 
was available for testing a code without the SSE optimizations was used. 
The Infiniband 
cluster performed approximately $1.8$ times better in the $32$-bit case and 
approximately 
$1.6$ times better in the $64$-bit case as the Myrinet cluster.\\
The $1$-dimensional (non-optimized) parallel even-odd preconditioned Dirac 
Operator Benchmark
was  modified to test whether an improvement using non-blocking MPI 
communication functions 
can be achieved. No effect was seen in using MPICH over GM provided by Myricom. 
Tests on a $4$ 
node cluster which runs the MPI version of the ParaStation software results in 
an performance 
increase of about $20$\% (see tab.~\ref{tab:nonblocking}).\\ 
\begin{table}[ht]
\begin{center}
\caption{Parastation3 non-blocking I/O support (non-SSE).\label{tab:nonblocking}}
\begin{tabular}{|l|l|}
\hline
\textbf{MPI blocking I/O} & \textbf{MPI non-blocking I/O}\\
\hline
$308$\,Mflops & $367$\,Mflops\\
\hline
\end{tabular}
\end{center}
\end{table}
\par
\begin{figure}[ht]
\begin{center}
\includegraphics[width=65mm]{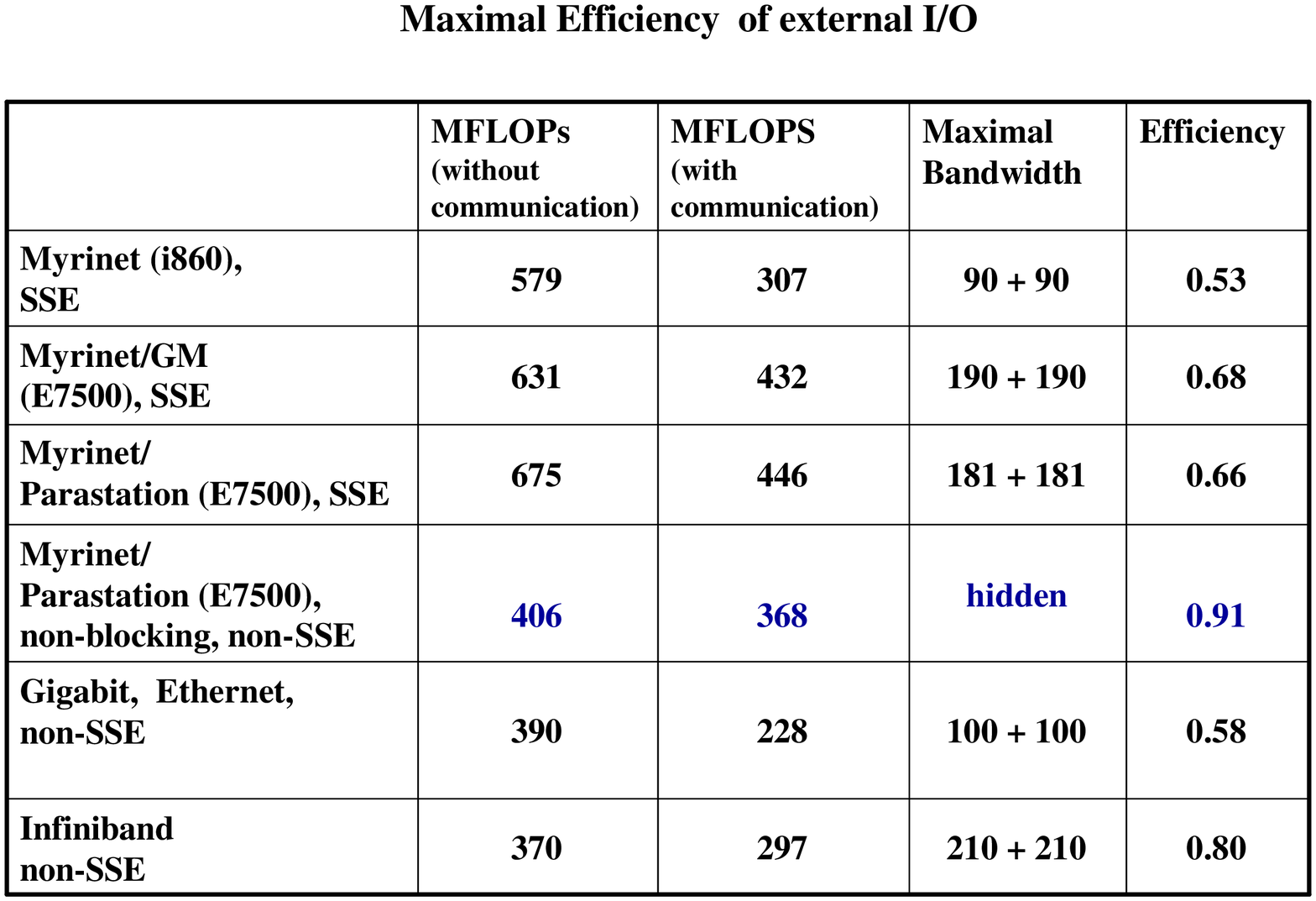}
\caption{I/O Efficiency.\label{fig:diagram6}}
\end{center}
\end{figure}

Fig.~\ref{fig:diagram6} gives a summary of the communication behavior of the 
different architectures resulting from the parallel benchmarks.
The efficiency number is the ratio between the  performance of the same
benchmark with and without communication.
Included is also a 
test on a $4$ node GigaBit-Ethernet cluster connected via a non-blocking 
GigaBit-Ethernet switch. The efficiency number for GigaBit-Ethernet in the 
rightmost column of fig. is ~\ref{fig:diagram6} even better than one can achieve 
using Myrinet on a system with a chip-set which provides a slow PCI-bus 
throughput.
The positive influence of non-blocking communication support (ParaStation) and 
fast communication support (Infiniband) is shown by the efficiency numbers.
\newline
Compared to the single node benchmarks the results of the parallel 
benchmarks imply that that the capacity of bandwidth is crucial for the 
efficient use of PC clusters as a  scalable platform for LQCD applications.
%
%
\section{Conclusions} \label{conclusions}
We have discussed the usage of commodity clusters based on PCs for LQCD 
calculations. Such installations would not have been considered
competitive a few years ago. However, our experience with such kind of
machines at NIC and DESY adds further evidence that for problems in
QCD that require below, say, 1 Tflops computer power, 
PC clusters are a valuable and cots-effective tool for computing
physics results in LQCD. 
In Hamburg and Zeuthen clusters with $64$ and $32$ CPUs are successfully 
in operation for more than one year.\\
Investigations using representative benchmarks on the DESY clusters and also
other architectures were carried out with promising results.\\ 
Applying the SSE/SSE2 (SIMD+pre-fetch) instructions on Pentium 4 like CPUs, the 
single node performance of the Wilson-Dirac operator is increasing according to
the 
clock rate improvements of those CPUs used in commodity PCs. \\
The performance of memory bounded parts of the LQCD algorithms, especially the 
linear algebra routines, depends strongly on the throughput of the memory 
interfaces.
Those interfaces did not show the same level of enhancements as it has been
observed and will be expected further in the case of the CPUs.\\
The performance of the $1$-dimensional and $2$-dimensional parallel 
implementations 
of the Dirac-Wilson operator depends on the behavior of the external 
interconnects,
i.e. is mainly dependent on the PCI-bus throughput coming given by the chip-set 
and the interface card itself.
Results coming from PC clusters consisting of 
different components have shown an enhancement in both the quality of the 
chip-sets (e.g. E7500)and the throughput of the communication interfaces
(e.g. Infiniband).\\
Non-blocking MPI communication can improve the performance by  
using adequate MPI implementations (e.g. ParaStation).
\par
In summary, it might be envisaged, as done by e.g. LATFOR,
that heterogeneous computer landscapes
will be available to the user with centers that host machines in the 
multi-Tflops regime, still enabled by specialized machines, 
and many smaller installations at universities as well as research centers
in the few hundred to 1 Tflops range realized by PC clusters.
%
%
\section*{Appendix: Discussion of the even-odd benchmark}\label{appendix}
The benchmark program applies the even-odd preconditioned Wilson-Dirac
matrix that is defined on a $L^3 \times T$ lattice to a spinor-field. 
The program is implemented in C plus some in-lined SSE2 extensions. 
For parallelization we have used the MPI message passing library.
The code is derived from Martin L\"uscher's benchmark code presented at 
the lattice conference 2001 \cite{Luscher2001}. 
Even with communication switched off, the present code performs worse 
then the one of Martin L\"uscher
($579$\,Mflops vs. $880$\,Mflops on $1.7$\,GHz Pentium~4). 
The reason is twofold:
\begin{itemize}
\item Less variables that reside in the cache can be reused than 
      in the standard case.
\item We have skipped the cache-optimized order 
      of the lattice-points to simplify the parallelization.
\end{itemize}
\subsection*{Strategy of the parallelization}
The Wilson-Dirac matrix is a sparse matrix. The hopping part of the matrix
only connects nearest neighbor sites on the lattice.  
Therefore, for parallelization, it is natural to divide the lattice 
in sub-blocks
of size $t \times l_x \times l_y \times l_z$. Each of the MPI-processes 
takes one such sub-block.  
For simplicity we have done the parallelization only in one direction:
$l_x=l_y=l_z=L$ and $t n_p = T$, where $n_p$ is the number of processes.
\par
The hopping part of the Wilson-Dirac matrix connects nearest neighbor sites
on the lattice. Therefore each application of $H_{eo}$ or $H_{oe}$
($H_{oe}$ connects even with odd sites and $H_{eo}$ vice versa.) 
the spinor-fields at the right boundary of the left neighbor and 
the spinor-fields at the left  boundary of the right neighbor 
of each of the processes has to be sent and received.
\par
In the case of even-odd pre-conditioning the spinor-field only resides 
on the even (or odd) sites. Therefore
$L^3/2$ spinors have to be sent and received.
A single
data package has the size
\begin{equation}
24 \times 8 \times L^3/2 \;\;\; \mbox{Byte} = 96 \times L^3 \;\;\; \mbox{Byte}
\end{equation}
\par
In our blocking implementation, 
the communication  of the data and the computation 
is performed in a consecutive way.  First the spinor-fields are exchanged
using the $\mbox{MPI}\_\mbox{Sendrecv}$ function. 
This is followed by the application 
of $H_{eo}$ or $H_{oe}$ on the single nodes. 
\par
Both times required for communication $t_{comm}$ and calculation $t_{calc}$ 
are measured separately. The effective bandwidth is computed as:
\begin{equation}
\mbox{Bandwidth} = \frac{96 \times L^3/2 \;\; \mbox{Byte}}{t_{comm}}
\end{equation}
The performance per node without communication is computed as:
\begin{equation}
P_0 = \frac{t \times L^3/2 \times 1392 \; \mbox{flop}}
                           {t_{calc}}
\end{equation}
Correspondingly, the performance including the communication of the data
is given by:
\begin{equation}
P = \frac{t \times L^3/2 \times 1392 \; \mbox{flop}}
                         {t_{comm}+t_{calc}}
\end{equation}
\par
In the case of the non-blocking communication, we had to divide $H_{oe}$ 
(or $H_{eo}$) into a part that only acts on spinors  that reside on the
local lattice and a part that  acts on spinors that reside on the neighbors:
\begin{itemize}
\item Initialize send and receive  ($\mbox{MPI}\_\mbox{Isend}$ and 
$\mbox{MPI}\_\mbox{Irecv}$),
\item perform the calculation for the local part of $H_{oe}$,
\item Wait for the communication to finish ($\mbox{MPI}\_\mbox{Wait}$),
\item do the rest of $H_{oe}$.
\end{itemize}
%
%
\begin{acknowledgments}
The authors would like to thank Martin L\"uscher (CERN) for the benchmark codes
and the fruitful discussions about PCs for LQCD, and
Isabel Campos Plasencia (Leibnitz-Rechenzentrum Munich),
Gerhard Wellein (Uni Erlangen),
Holger M\"uller (MEGWARE),
Norbert Eicker (ParTec),
Chris Eddington (Mellanox)
for the opportunity to run the benchmarks on their clusters.
\newline
The authors also wish to thank the computer centers of DESY Hamburg and Zeuthen. 
\end{acknowledgments}
%
%

%
\end{document}